\documentstyle[12pt,epsfig]{article}
\begin{document}
\title{\vspace{-15mm}
       {\normalsize \hfill
       \begin{tabbing}
       \`\begin{tabular}{l}
	 SLAC--PUB--6506 \\
	 May 1994 \\
	 hep--th/9407039 \\
	 T   \\
	\end{tabular}
       \end{tabbing} }
       \vspace{8mm}
\setcounter{footnote}{1}
Towards Quantum Cosmology without Singularities\thanks{Work
supported by the Department of Energy, contract DE--AC03--76SF00515.}}

\renewcommand{\thefootnote}{\fnsymbol{footnote}}
\vspace{15mm}

\author{
\setcounter{footnote}{2}
Klaus Behrndt\thanks{e-mail: behrndt@jupiter.slac.stanford.edu} 
\qquad and \qquad 
Thomas T. Burwick\thanks{e-mail: burwick@slacvm.slac.stanford.edu}
\\[2mm] {\normalsize \em
 Stanford Linear Accelerator Center} \\
{\normalsize \em Stanford University, Stanford, California 94309}
}
\date{}
\maketitle
\renewcommand{\arraystretch}{2.0}
\renewcommand{\thefootnote}{\alph{footnote}}
\newcommand{\be}{\begin{equation}}
\newcommand{\ee}{\end{equation}}
\newcommand{\ba}{\begin{array}}
\newcommand{\ea}{\end{array}}
\newcommand{\vsf}{\vspace{5mm}}
\newcommand{\NP}[3]{{\em Nucl. Phys.}{ \bf B#1#2#3}}
\newcommand{\PRD}[2]{{\em Phys. Rev.}{ \bf D#1#2}}
\newcommand{\MPLA}[1]{{\em Mod. Phys. Lett.}{ \bf A#1}}
\newcommand{\PL}[3]{{\em Phys. Lett.}{ \bf B#1#2#3}}
\newcommand{\marpar}{\marginpar[!!!]{!!!}}
%\vfill
\begin{abstract}
In this paper we investigate the vanishing of cosmological
singularities by quantization. Starting from a 5d Kaluza--Klein
approach we quantize, as a first step, the non--spherical
metric part and the dilaton field. These fields which are 
classically singular become smooth after quantization.
In addition, we argue that the incorporation of 
non perturbative quantum corrections form a dilaton potential.
Technically, the procedure corresponds to the quantization
of 2d dilaton gravity and we discuss several models.
From the 4d point of view this procedure is a semiclassical
approach where only the dilaton and moduli
matter fields are quantized.
\end{abstract}
%\thispagestyle{empty}
%\vfill
%\vspace{10mm}
%\begin{center}
%Submitted to {\em Phys. Rev. Lett. }
%\end{center}
%\vfill
%\newpage
\noindent
We consider a cosmological string solution which has classical
singularities (big bang). Near these singularities the theory
factorizes in a smooth spherical part and a singular 2d part.  This
singular part is the well-known dilaton gravity (see e.g. 
\cite{cghs}-\cite{sussk}) and as a first step we 
are going to quantize this part with
the result that all singularities disappear. This procedure is also
known as s-wave reduction and was so far used for 4d BH physics.

\vspace{5mm}

\noindent
{\bf 1. Classical theory}. Our 4d classical model is given by
\be  \label{1}
S = \int d^4x \sqrt{\tilde{G}} e^{-2 \phi}
 \left( R + 4 (\partial\phi)^2 -
(\frac{\partial \rho}{\rho})^2 - \frac{1}{12} H^2 \right) 
\ee
where $\phi$ is a dilaton field, $H_{\mu\nu\lambda} =
\partial_{[\mu} B_{\nu\lambda]}$ is the torsion corresponding to the
antisymmetric tensor field $B_{\mu\nu}$ and $\rho$ is a modulus field.
A cosmological solution to this model is given by \cite{behr}
\begin{equation} \label{2}
\begin{array}{l c l}
ds^2  =  -\frac{dt^2}{\left( -k+\left(\frac{t_+}{t}\right)^2\right)
\left( 1-\left(\frac{t_-}{t}\right)^2\right)}+t^2d\Omega_k^2 & , &
\rho^2  = \frac{-k+\left(\frac{t_+}{t}\right)^2}{1-\left(\frac{t_-}{t}
\right)^2} \\
H = 2 t_+ t_- \left( \frac{\sin \sqrt{k} \chi}{\sqrt{k}}\right)^2
\sin \theta \, d\chi \wedge d\theta \wedge d\varphi & , &
e^{-2\phi}  \sim  \sqrt{ \left( 1-\left(\frac{t_-}{t}\right)^2\right)
\left( -k+\left(\frac{t_+}{t}\right)^2\right)} \ .
\end{array} \end{equation}
After a time reparameterization one obtains the standard
Friedmann--Robertson--Walker (FRW) metric with $d \Omega_k^2$ as 3d
volume form corresponding to the spatial curvature $k$
($=0,-1,+1$). The parameter $t_-$ is the minimal extension and for
$k=1$ the parameter $t_+$ denotes the maximal extension of the
universe. This is obvious after transforming the solution to the
conformal time
\be \label{3}
t^2 = t_{-}^2 + ( t_{+}^2 - k t_{-}^2) \left(\frac{\sin \sqrt{k} \eta}
 {\sqrt{k}}\right)^2  
\ee
for which the metric is
\be \label{4}
ds^2 = \left\{ t_{-}^2 + (t_{+}^2 - k t_{-}^2) \left(\frac{\sin \sqrt{k}
  \eta }{\sqrt{k}}\right)^2 \right\} \, \left[ -d\eta^2 +
  d\Omega_{k}^2 \right] \ .
\ee
\begin{figure}[t] \vspace*{1mm}
%\begin{minipage}[t]{155mm}
\hspace{-10mm}
\begin{tabular}{lr}
 \begin{minipage}[t]{8cm}
\mbox{\epsfig{file=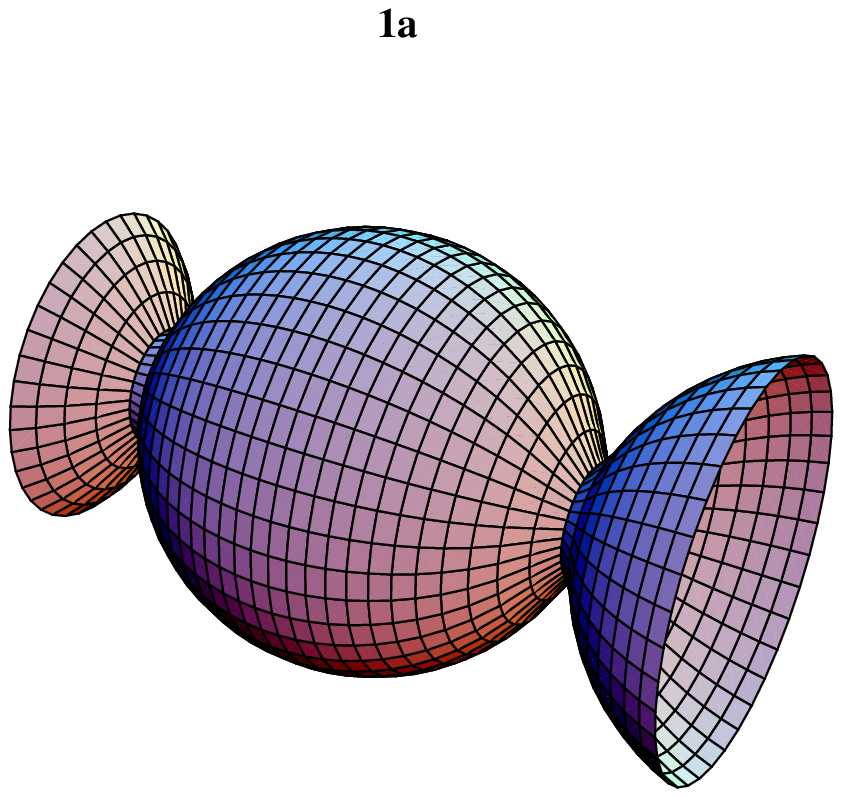,height=7cm,width=7cm}}
     \end{minipage}
 \hspace{5mm} &
  \begin{minipage}[t]{8cm}
 \mbox{\epsfig{file=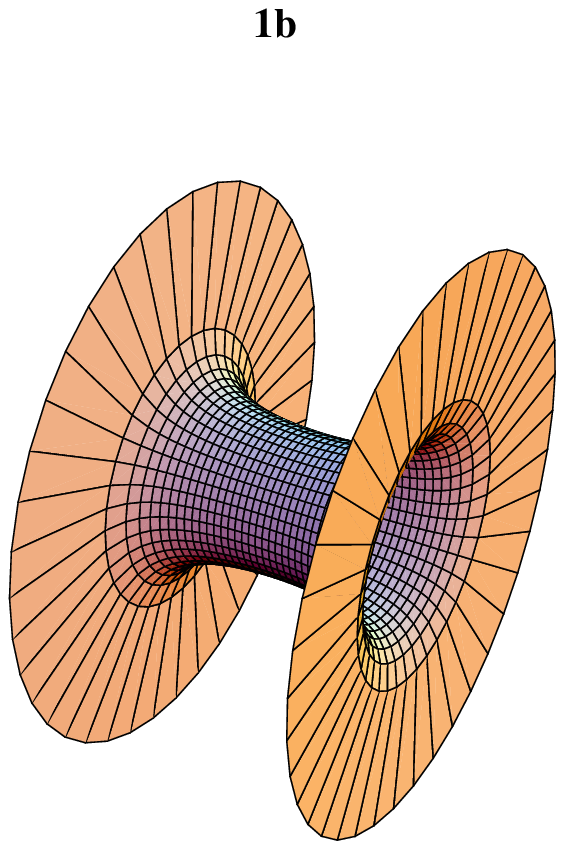,height=7cm,width=7cm}}
 \end{minipage}
\end{tabular} \vspace{-10mm}
\caption{In (a) we have plotted the closed oscillating solution
for $k=1$ and (b) is the wormhole solution for $k=-1$.}
%\end{minipage}
\vspace{2mm}
\end{figure}
Unfortunately, in \cite{behr} no analytic results for the world radius
$a(\tau)$ in the standard parameterization of the FRW metric
$ds^2=-d\tau^2 + a(\tau)^2 d\Omega_k^2$ could be found. We have plotted
numerical results in figure 1. Figure 1a shows the oscillating
solution for $k=1$ and 1b the wormhole solution for $k=-1$.  For $k=0$
the solution has again the geometry of figure 1b but with the
difference that there are no asymptotic flat regions as for $k=-1$.
Remarkably, the scalar fields $\rho$ and $\phi$ have divergencies
although the metric behaves completely smooth for all times. To
understand this phenomena one has to go back to the 5 dimensional (5d)
origin. In the sense of a Kaluza--Klein approach solution (2)
can be obtained by dimensional reduction of a 5d theory. Then, the
modulus field $\rho$ corresponds to a time dependent compactification
radius of the fifth coordinate. The corresponding action is given
by the effective string action
\be   \label{5}
S^{(5)} = \int d^5 x \sqrt{G} e^{-2 \psi} \left( R + 4 (\partial \psi)^2
  - \frac{1}{12} H^2 \right)
\ee
and the 5 dimensional solution can be written as
\be \label{6}
\ba{l}
ds^2 = \left( \frac{\sqrt{k}}{\tan \sqrt{k} \eta } \right)^2 \, dw^2
  + \left\{ t_{-}^2 + (t_{+}^2 - k t_{-}^2) \left(\frac{\sin \sqrt{k}
  \eta }{\sqrt{k}}\right)^2 \right\} \, \left[ -d\eta^2 +
  d\Omega_{k}^2 \right] \\
 e^{2 (\psi - \psi_0)} = 1 + \frac{t_{-}^2}{(t_{+}^2 - k t_{-}^2)
 \left(\frac{\sin \sqrt{k} \eta }{\sqrt{k}}\right)^2} \ .
\ea
\ee
The 5d dilaton is related to $\phi$ by $2 \psi=2 \phi + \log \chi$
and $\rho = G_{55}$. For $k=1$ and after switching the signature of
the metric ($dw^2 \rightarrow -dw^2$ and $d\eta^2 \rightarrow
-d\eta^2$) this 5d solution is just the 5d BH solution (in the
conformal time $\eta$) discussed by Horowitz and Strominger
in \cite{strom2}. Here, $t_{\pm}$ define the two horizons of the
theory and our cosmological solution lives between these horizons.

\vspace{5mm}

\noindent
{\bf 2. S-wave reduction}. We are particularly interested in the fate of
the singularities if one starts to quantize the theory.  Therefore, it
is reasonable to restrict ourselves to the region near the
singularities ($\sin \sqrt{k} \eta\simeq 0$). In this region one can
assume that quantum corrections become important. Furthermore, as one
can see from (\ref{6}) in this limit the 5d solution decouples in a
3d spherical part ($\sim d\Omega_k^2$) and a 2d $(w,\eta)$ part which
is the known dual 2d BH \cite{witt}. In the figure it is just the region of
minimal extension, e.g.\ inside the wormhole of fig.\ 1b.

Before we can start to quantize the 2d part we have to reduce the 5d
action (\ref{5}) down to a 2d theory. Motivated is this procedure by
the assumption that the quantum corrections respect the spherical
symmetry. Generally, this is not the case but sufficient for a first 
approximation. In BH physics this procedure is also
known as s--wave reduction. For the 5d metric we make the ansatz
\be \label{7}
ds^2 = g^{(2)}_{ab} dz^a dz^b + e^{2 \chi} d\Omega_k^2
\ee
where $g^{(2)}_{ab}$ is the 2d metric part. In what follows we
quantize only $g^{(2)}_{ab}$ and the dilaton $\psi$ (resp. $\phi$, see
below). The remaining fields are assumed to be classical backgrounds
given by (2) or (6).  After integrating out the angular degrees of
freedom and using the $H$ field from (\ref{2}) we obtain for (\ref{5})
\be \label{8}
S^{(2)} = \int d^2z \sqrt{g} e^{-2 \phi} \left( R^{(2)} + 4 (\partial \phi)^2
 - 3 ( \partial \chi)^2 + V(\chi) \right) 
\ee
with $\phi = \psi + \frac{3}{2} \chi$ and $ V(\chi) = 6 k e^{-2 \chi}
- 2 t_+^2 t_-^2 e^{-6 \chi}$.  Near the singularity ($\eta \simeq 0$)
the background field $\chi$ is smooth, $\frac{\partial}{\partial \eta}
\chi |_0 = 0$ (see (6) and (\ref{7})) and up to the second order in 
$\eta$ we can approximate the $\chi$ terms by a constant
\be \label{9}
S^{(2)} = \int d^2z \sqrt{g} e^{-2 \phi} \left( R^{(2)} + 4 (\partial
  \phi)^2 + \lambda \right)
\ee
with $\lambda = \frac{2}{t_-^2} \left(3k - (\frac{t_+}{t_-})^2 \right)$.
A classical solution in conformal coordinates is given by \cite{witt}
\be  \label{91}
ds^2 = e^{2 \sigma} dz^+ dz^-  \qquad , \qquad e^{-2 \phi} \sim e^{-2 \sigma}
= u - \lambda z^+ z^-
\ee
where $u$ is constant. This solution can be transformed to the 
2d ($w,\eta$) part of (\ref{6}) where $\eta \simeq 0$ corresponds to
$u \simeq \lambda z^+ z^-$.

\vspace{5mm}

\noindent {\bf 3. Quantization}. The quantization of (9) has
been studied in various papers \cite{cghs}-\cite{sussk}, 
\cite{dealwis, bilal}. 
Each of these models will be discussed, but before we do so in
detail let us come back to the classical solution once more. The
fact that (6) or (2) are independent of tha fifth coordinate leads
to the dual solution
\be \label{10}
\ba{l}
ds^2 = \left( \frac{\tan \sqrt{k} \eta }{\sqrt{k}} \right)^2 \, dw^2
  + \left\{ t_{-}^2 + (t_{+}^2 - k t_{-}^2) \left(\frac{\sin \sqrt{k}
  \eta }{\sqrt{k}}\right)^2 \right\} \, \left[ -d\eta^2 +
  d\Omega_{k}^2 \right] \\
 e^{2 (\psi - \psi_0)} = \left(\frac{\tan \sqrt{k} \eta }{\sqrt{k}}\right)^2  
 + \frac{t_{-}^2}{(t_{+}^2 - k t_{-}^2)
 \left(\cos \sqrt{k} \eta \right)^2} \ .
\ea
\ee
Both solutions have a significant difference.  Singular points of
(\ref{6}) are regular in (\ref{10}) and vice versa. Furthermore, in
the region that we are interested in ($\sin\sqrt{k} \eta \simeq 0$)
the solution (\ref{6}) is in the strong coupling region ($e^{2 \psi}
\rightarrow \infty$) whereas the dual solution (\ref{10}) is in the
weak coupling region ($e^{2 \psi} << 1$; if we assume that $t_+>>t_-$
which is reasonable, since $t_{\pm}$ corresponds to the
maximal/minimal extension of the universe)
Again the 2d part decouples and can be transformed into (\ref{91})
where $\eta \simeq 0$ corresponds to $z^+ z^- \simeq 0$.

When quantizating this theory we are especially interested in what
happens in the strong coupling region, i.e. the fate of the
singularity in (\ref{6}). As a consistency condition of this procedure
we have to ensure that in the classical limit (weak coupling region)
we get back our classical result (\ref{10}) which is non singular for
$\eta \simeq 0$. We are following the procedure of de Alwis
\cite{dealwis} and later on we discuss the modification concerning the
other models.  After choosing the conformal gauge
\be  \label{11}  
g_{ab} = e^{2 \sigma} \hat{g}_{ab}
\ee
we can rewrite (\ref{9}) as a general 2d $\sigma$ model
\be \label{12}
S = -\int d^2 z \sqrt{\hat{g}} \left[ \hat{g}^{ab}\partial_a X^{\mu}
  \partial_b X^{\nu} G_{\mu\nu}(X) + \hat{R}\, \Phi(X) + T(X) \right]
\ee
with: $X^{\mu}=\{\phi,\sigma\}$. Thus, the quantization of the dilaton
gravity is reduced to the quantization of a 2d $\sigma$ model with the
target space spanned by $\phi$ and $\sigma$.  This model, however, is
well defined only if the background fields $G_{\mu\nu}, \Phi$ and $T$
define a 2d conformal field theory. This symmetry is a consequence of
the fact, that the original theory depends only on $g$ and not on
$\hat{g}$, and thus, has to respect the symmetry: $\hat{g} \rightarrow
e^{2 \rho} \hat{g}$ and $\sigma \rightarrow \sigma - \rho$ (see
(\ref{11})). We transform the theory to an exact model and  define 
the quantum theory by this (exact) conformal field theory (see e.g.
\cite{dealwis, bilal}). Following this
approach we first note that the target space metric $G_{\mu\nu}$ has
the general structure
\be  \label{13}
dS^2 = -4 e^{-2\phi} [1+h(\phi)] d\phi^2 + 4 e^{-2 \phi} [1+\bar{h}(\phi)]
d\sigma d\phi +  \kappa d\sigma^2 
\ee
where $h$ and $\hat{h}$ are model dependent functions of $\phi$ or
$X^1$. For $h=\bar{h}=0$ we have the CGHS model \cite{cghs}; for $ 2 h
= \bar{h} = -e^{2\phi}$ the model from Strominger \cite{strom}; $h=0$
and $\bar{h} = - \frac{\kappa} {4} e^{2 \phi}$ describes the RST model
\cite{sussk}. The parameter $\kappa=\frac{24-N}{6}$ originates from the
definition of the functional integration measure and $N$ corresponds to
additional conformal matter. As next step we introduce new target space
coordinates
\be  \label{14}
\ba{l}
x = \frac{2}{\sqrt{\kappa}} \int d\phi\, e^{-2 \phi} \sqrt{(1+\bar{h})^2
  +\kappa e^{2 \phi} (1+h)} \\
y = - \sqrt{\kappa} \left( \sigma - \frac{1}{\kappa} e^{-2 \phi} + \frac{2}
{\kappa} \int d\phi e^{-2 \phi} \bar{h} \right)
\ea
\ee
and obtain a flat metric 
\be \label{15}
dS^2 = - dx^2 + dy^2 \ .
\ee
(for negative $\kappa$ we have to perform a Wick rotation in $x$ and
$y$).  For this flat metric it is easy to find the dilaton $\Phi$ and
tachyon $T$ that define a conformal field theory. The general solution
of the corresponding $\beta$ equations is
\be  \label{16}
\ba{ll}
\Phi(x) = a x + b y & \qquad \mbox{with} \qquad a^2 - b^2 = 
 - \kappa \qquad ,\\
T(x) \sim e^{\alpha x + \beta y}  & \qquad \mbox{with} \qquad
    \frac{1}{2} (\alpha^2  - \beta^2) -  a \alpha +
   b \beta - 2 = 0 \ .
\ea
\ee
The demand to get the classical model (\ref{9}) in the weak coupling
limit yields a further restriction to $\Phi$ and $T$.  Following the
suggestion of de Alwis we set: $a=0$ and $\alpha = -\beta = -
\frac{2}{\sqrt{\kappa}}$ and get the known Liouville theory
($y$ as Liouville field) that couples to the matter field $x$
\be
S = - \int d^2 z \sqrt{\hat{g}} \left[- (\partial x)^2 + (\partial y)^2
 + \sqrt{\kappa} \, \hat{R} \, y + \lambda \, e^{-\frac{2}{\sqrt{\kappa}}
(x-y)} \right] \ .
\ee
This describes a well defined 2d gravity theory on the classical as
well as on the quantum level. The strategy is to define the
quantum theory in terms of {\em this} action and to regard
(\ref{9}) as the classical limit.

As second step we have to find solutions of the equations of
motion for $x$ and $y$ ($\hat{R}^{(2)}=0$)
\be  \label{17}
-  \partial^2 x = \frac{\lambda}{\sqrt{\kappa}} e^{-\frac{2}
{\sqrt{\kappa}} (x - y)} \qquad , \qquad \partial^2 y = \partial^2 x \ .
\ee
Solving these equations we have to restrict ourselves to solutions
that reproduce the BH solution (\ref{91}) in the classical limit.
Therefore, we are interested in a solution depending on $z^+ z^-$ only
and find
\be  \label{18}
x = y = - \frac{1}{\sqrt{\kappa}} \left(u - \lambda z^+ z^- \right)
\ee
($u=const.$). Using the transformation (\ref{14}) we can express this
solution in $\phi$ and $\sigma$. In doing so we have to fix the up to
now arbitrary functions $h(\phi)$ and $\bar{h}(\phi)$.  Let us start
with parameterization suggested by de Alwis: $h=0$,
$\bar{h}=-\frac{1}{2} \kappa e^{2 \phi}$.  Motivated is this choice by
the fact that for all values of $\phi$ and $\sigma$ the transformation
(\ref{14}) is non singular and secondly that the range of $x$ and $y$
goes from $-\infty$ to $+\infty$ if $\phi$ and $\sigma$ do so. 
For $x$ and $y$ one gets
\be  \label{19}
\ba{l}
x = \frac{1}{\sqrt{4 \kappa}}\left( - \sqrt{\kappa^2 + 4 e^{-4 \phi}}
  + \sqrt{\kappa}\, \mbox{arcsinh} \frac{\kappa}{2} e^{2\phi} \right) \\
y = - \sqrt{\kappa} \left( \sigma - \frac{1}{\kappa} e^{-2 \phi}
  - \phi \right) 
\ea 
\ee 
In terms of (\ref{18}) one finds in the weak coupling limit ($e^{2
\phi} << 1$) the desired classical solution (\ref{91})
\be  \label{21}
e^{-2\phi} = u - \lambda z^+ z^- \qquad , \qquad \sigma = \phi\ .
\ee
Since we are in the weak coupling region this solution corresponds to
our dual solution (11) which is non-singular
for $\eta \simeq 0$. In the
strong coupling limit ($e^{2 \phi} >> 1$) we obtain
\be  \label{22}
\phi = - \frac{1}{\sqrt{\kappa}} ( u - \lambda z^+ z^-) \qquad , 
 \qquad \sigma = \frac{1}{\kappa} e^{-2 \phi} \ .
\ee
Therefore, after incorporation of quantum corrections ($\sim
\cal{O}(e^{2 \phi})$) the black hole solution gets smooth also in the
strong coupling region. Note, that in dilaton gravity a
singularity in the metric has to be accompanied by a singularity in
the dilaton, i.e.\ singularities can only appear in the strong or weak
coupling region.  For the other models the picture is qualitatively the
same. In the CGHS model ($h=\bar{h}=0$) one obtains for (\ref{14})
\be  \label{23}
\ba{l}
x = -\frac{1}{\sqrt{\kappa}} e^{-2 \phi} \sqrt{1 + \kappa e^{2 \phi}} 
  - \frac{\sqrt{\kappa}}{2} \log \left[ \kappa + 2 e^{-2 \phi} (1 + 
\sqrt{1+\kappa e^{2 \phi}})\right]   \\
y = - \sqrt{\kappa} \left(\sigma - \frac{1}{\kappa} e^{-2 \phi} \right) \ .
\ea
\ee
and in strong coupling region this model gives
\be
e^{ -\phi} \sim u -\lambda z^+ z^- \qquad , \qquad \sigma = 
\frac{1}{\kappa}(u -\lambda z^+ z^-)
\ee
For the Strominger model ($2 h = \bar{h} = - e^{2 \phi}$) we find
($F(\phi) = \sqrt{e^{-4 \phi} - (2-\kappa) e^{-2 \phi} + 
\frac{2-\kappa}{2}} $\ )
\be  \label{24}
\ba{l}
x = -\frac{1}{\sqrt{\kappa}} \left[F(\phi) + \frac{\kappa -2}{2}
 \log[F(\phi) + e^{-2 \phi} + \frac{\kappa -2}{2}] - \right. \\
\qquad \qquad - \left. \sqrt{\frac{2-\kappa}{2}}
 \log\left(\sqrt{2(2-\kappa)} F(\phi) + (2-\kappa) e^{2 \phi} - 
  (2-\kappa)\right) \right]
\\
y = - \sqrt{\kappa} \left(\sigma - \frac{1}{\kappa} e^{-2 \phi} - 
  \frac{2}{\kappa} \phi \right) \ .
\ea
\ee
which gives in the strong coupling region
\be  \label{241}
\phi = - \frac{1}{\sqrt{2(2-\kappa)}}(u -\lambda z^+ z^-) 
\qquad , \qquad \sigma = \frac{1}{\kappa} (\sqrt{2(2-\kappa)} -2) \phi \ .
\ee
And finally in the RST model ($h=0$, $\bar{h}=-\frac{\kappa}{4} e^{2 \phi}$)  
the general solution is given by
\be   \label{25}
\ba{l}
x = - \frac{1}{\kappa} e^{-2 \phi} + \frac{\kappa}{2} \phi \\
y = - \sqrt{\kappa} \left(\sigma - \frac{1}{\kappa} e^{-2 \phi} - 
  \frac{1}{2} \phi \right) \ .
\ea
\ee
In the strong coupling region this model behaves like
\be   \label{251}
\phi = - 2 \kappa^{- \frac{3}{2}}(u -\lambda z^+ z^-) 
\qquad , \qquad \sigma = \frac{1 - \sqrt{\kappa}}{2} \phi \ .
\ee
Therefore we find that all models have no singularities in the strong coupling
region and yield the classical result in the weak coupling region.

One can now ask what is the influence of this quantization procedure
for the further evolution of the universe. For the derivation of our
results it was crucial that the solution decouples in a 2d (dilaton
gravity) part and a 3d spherical part. This is valid only if one
considers the theory, e.g.\ ,inside the wormhole of fig.\ 1b. Extending
this procedure to the region away from the wormhole seems
to be difficult. But nevertheless, quantum correction inside the
wormhole can form a dilaton potential which could be a source of an
inflationary period in later times.  A dilaton potential in our
original action (\ref{1}) or (\ref{5}) corresponds to
an additional tachyon contribution in the 2d action which is independent 
of $\lambda$
(since $\lambda$ was correlated to the constant $\chi$ field in the
wormhole, see (\ref{8})). The tachyon we have discussed so far is only
{\em one} possibility.  This solution has the advantage that the
renormalization group $\beta$ functions vanish thereby yielding a
finite 2d quantum field theory. The most general tachyon field,
however, is a combination of contributions given by (17).  A
further additive tachyon term is given by (for $\kappa > 0$)
\be \label{31}
T_{np} = \mu e^{2 x} 
\ee
where the function $x$ is given by (\ref{14}) This term, discussed
e.g.\ in \cite{dealwis} and \cite{tseyt}, has in the weak coupling
region for all discussed models the typical non-perturbative structure
\be
T_{np} \sim e^{-\frac{2}{\sqrt{\kappa}} e^{-2\phi}} \sim 
e^{- \frac{2}{\sqrt{\kappa} g^2_s}}
\ee
where $g_s = e^{\phi}$ is the string coupling constant. Therefore this
term vanishes very rapidly in the weak coupling (classical) region
and becomes important in the strong coupling region. Furthermore,
since $x$ is a function of the dilaton only,
this tachyon term represents a candidate for a dilaton potential
created by non-perturbative quantum corrections in the strong coupling
region. If we insert the $x$ values for the several models (21),
(\ref{23}), (\ref{24}) and (\ref{25}) we obtain different potentials.
But all these potentials have no local or global minima and are
probably not good candidates to discuss an inflationary period (see
e.g.\ \cite{ramy} and refs.\ therein).  It remains an open question
whether another choice of the model dependent functions $h$ and
$\bar{h}$ could yield a more appropriate potential.

\vspace{5mm}

\noindent
{\bf 4. Discussion}. Starting with a classical solution of the low
energy string effective action we investigated the quantization near
the cosmological singularity. Via a 5d Kaluza--Klein approach this
solution was obtained as a dimensional reduced theory.  Near the
singularity the 5d theory decouples in a 3d non singular (spherical)
part and a singular 2d part. As a first step we have quantized only
this singular 2d part (s-wave reduction).  The results (\ref{19}) -
(\ref{251}) show that for all models the singularity disappears after
the quantization of the theory, i.e. the 2d metric part and the
dilaton remains finite. An interpretation of this result is that the
wormhole becomes traversable via quantum corrections. In addition, we
have shown that the incorporation of non perturbative quantum
corrections form a dilaton potential.  The discussion of the possible
structures of the potential created by this procedure remains an
interesting task for further investigations.

We used the 5d theory to get contact
with the known dilaton gravity. But it is also possible to quantize
the 4d theory (\ref{1}) directly.  Our approximation to quantize
only the divergent 2d part in 5 dimensions is effectively the same as
to quantize the dilaton and moduli matter fields only. Note that the
2d metric part has only one degree of freedom. In the conformal gauge
(\ref{11}) this is the Liouville field $\sigma$ but we can also take
another gauge, e.g. $ds^2 = \rho^2 dw^2 - d\eta^2$ and then $\rho$ is
our moduli field (see (\ref{6})). Thus, from the 4d point of view 
we replaced the dilaton and moduli contributions in the Einstein equation  
by its vacuum expectation value
\be
R_{\mu\nu}^{(E)} - \frac{1}{2} R^{(E)} G_{\mu\nu}^{(E)}\  = \  
<T_{\mu\nu}^{(\phi, \rho)}> + \, T_{\mu\nu}^{(H)}
\ee
where $G^{(E)} = e^{-2 \phi} G_{\mu\nu}$ is the metric in the Einstein frame.
Classically, the 4d string metric was smooth but the Einstein metric was
singular (caused by the dilaton and moduli). However, after quantization
the singularities in the scalar fields disappeared and thereby also
the Einstein metric turned out to be non-singular. This implies that
similar to the string frame the Einstein metric describes a universe
which starts and ends (for $k=1$) {\bf not} with a singularity but
with a minimal (nonzero) extension (wormhole). Therefore, in both
frames the spatial part of the universe is qualitatively given in figure
1. Of course, the quantization of the scalar fields near the
singularity can only be a first step and future investigations have to
show whether a complete quantum theory will leave this qualitative 
feature intact.

\vspace{5mm}

\noindent
{\large\bf Acknowledgments}\vspace{3mm}\newline\noindent
We would like to thank S. F\"orste and J. Garc\'{\i}a-Bellido for
useful comments.  The work of K.B. is supported by a DAAD grant and of
T.T.B. by the Swiss National Science Foundation.  The work of both
authors was also supported by the US Department of Energy, contract
DE-AC03-76SF00515.

\end{document}